\documentclass[aps,prd]{revtex4}
\usepackage{epsf,epsfig,graphicx}
\begin{document}

\title{Comment on ``Gauge Invariance and $k_T$-Factorization of Exclusive Processes"}

\author{Hsiang-nan Li$^1$}
\email{hnli@phys.sinica.edu.tw}
\author{Satoshi Mishima$^2$}
\email{satoshi.mishima@desy.de}

\affiliation{$^1$Institute of Physics, Academia Sinica, Taipei,
Taiwan 115, Republic of China,}

\affiliation{$^1$Department of Physics, National Cheng-Kung
University, Tainan, Taiwan 701, Republic of China}

\affiliation{$^1$Department of Physics, National Tsing-Hua
University, Hsinchu, Taiwan 300, Republic of China}

\affiliation{$^2$Theory Group, Deutsches Elektronen-Synchrotron
DESY, 22607 Hamburg, Germany}

\begin{abstract}

We point out mistakes made in the one-loop calculation of some
diagrams for the process $\pi \gamma^* \to \gamma$ in the preprint
arXiv:0807.0296, and present correct results. Especially, we have
difficulty to understand their argument that a highly off-shell
gluon generates a light-cone (infrared) singularity. It is shown
by means of the Ward identity that the gauge-dependent light-cone
singularity found in arXiv:0807.0296 does not exist. It is then
shown that a hard kernel derived in the $k_T$ factorization of
exclusive processes is gauge invariant and free of the light-cone
singularity.

\end{abstract}


\maketitle

In a recent preprint arXiv:0807.0296 \cite{FMW08}, the authors
studied the pion transition form factor in the $k_T$ factorization
theorem at one-loop level, calculating the diagrams in full QCD for
the form factor (see their Fig.~4) and the diagrams for the
$k_T$-dependent pion wave function (see their Figs.~2 and 3) in the
covariant gauge. They focused on a special piece of contribution,
which depends on the gauge parameter and contains a light-cone
singularity. It was found that the full QCD diagrams do not generate
this contribution, but those for the pion wave function do, if the
involved partons are taken to be off-shell. Therefore, the hard
kernel, defined as the difference of the above two sets of diagrams,
is gauge dependent, and contains the light-cone singularity. They
were then led to the conclusion that the $k_T$ factorization theorem
for exclusive processes violates gauge invariance, and that the
perturbative QCD (PQCD) approach \cite{LS,LY1,KLS,LUY} to exclusive
$B$ meson decays, based on the $k_T$ factorization theorem, is also
gauge dependent. This contradicts the conclusions drawn in our
previous work \cite{NL07}. In this comment we point out the
mistakes made in their calculation, and demonstrate that the
gauge-dependent light-cone singularity discussed by the authors of
\cite{FMW08} does not exist. The all-order proof for the gauge
invariance and infrared finiteness of a hard kernel derived in the
$k_T$ factorization \cite{NL07} is indeed correct.

\begin{figure}[t]
\begin{center}
\includegraphics[height=3.5cm]{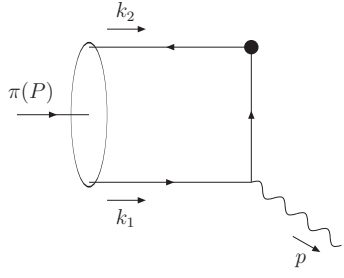}
\caption{LO diagram for $\pi(P) \gamma^* \to \gamma(p)$.}
\label{fig1}
\end{center}
\end{figure}

\begin{figure}[t]
\begin{center}
\includegraphics[height=3cm]{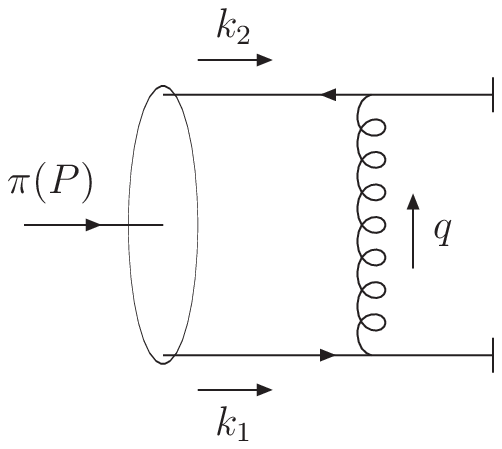}
\includegraphics[height=3cm]{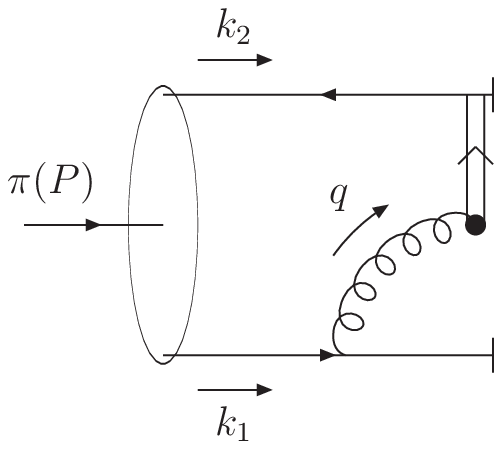}
\includegraphics[height=3cm]{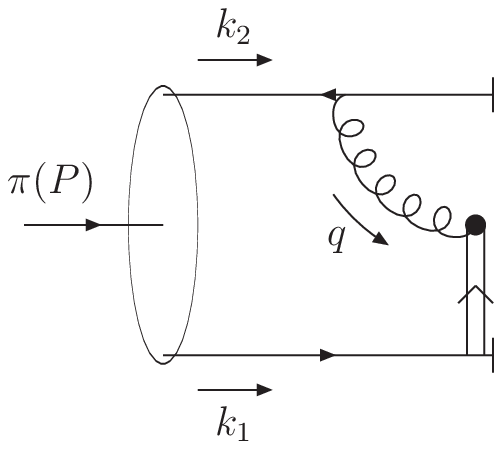}
\includegraphics[height=3cm]{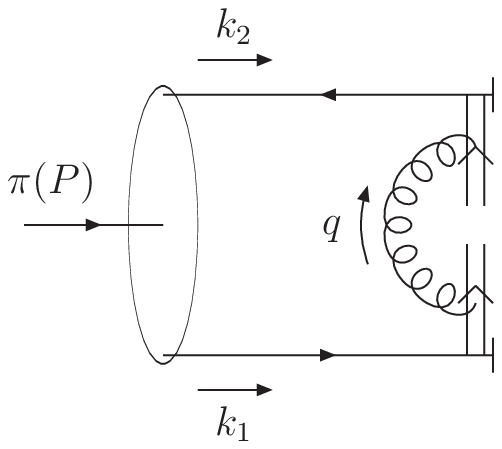}

\hspace{1.0cm} (a)\hspace{3.0cm} (b) \hspace{3.0cm} (c)
\hspace{3.0cm} (d)

\caption{NLO diagrams for the pion wave function with the loop
momentum flowing through the hard kernel.} \label{fig2}
\end{center}
\end{figure}

Consider the process $\pi(P) \gamma^* \to \gamma(p)$ in
Fig.~\ref{fig1}, with $P$ ($p$) being the pion (outgoing on-shell
photon) momentum along the plus (minus) direction. Define the
momenta of the valence quark and anti-quark in the pion as
\cite{FMW08}
\begin{equation}
k_1^\mu =(k_1^+, 0,\vec k_{1\perp})\;, \ \ \ \ k_2^\mu =(k_2^+, 0,
-\vec k_{1\perp})\;,
\end{equation}
respectively, with $k_1^+=x_0P^+$ and $k_2^+=(1-x_0)P^+$, $x_0$
being the momentum fraction. The leading-order (LO) hard kernel is
given, in terms of the above momenta, by
\begin{eqnarray}
H^{(0)}=\frac{1}{2k_1^+p^-+k_{1\perp}^2}
=\frac{1}{x_0Q^2+k_{1\perp}^2}\;,\label{1}
\end{eqnarray}
with the momentum transfer $Q^2=2P^+p^-$. At next-to-leading order
(NLO), a loop momentum $q$ carried by the additional gluon may flow
through the hard kernel, for example, in Fig.~\ref{fig2}. In this
case the factorized $k_T$-dependent wave function is convoluted with
\begin{eqnarray}
H^{(0)}=\frac{1}{2(k_1^+-q^+)p^-+|\vec k_{1\perp}-\vec
q_\perp|^2}\;.\label{2}
\end{eqnarray}

\begin{figure}[t]
\begin{center}
\includegraphics[height=3cm]{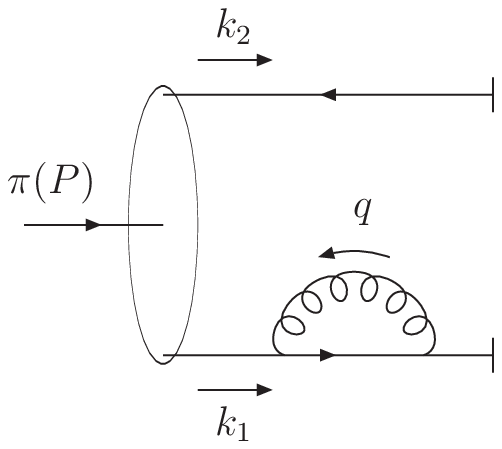}
\includegraphics[height=3cm]{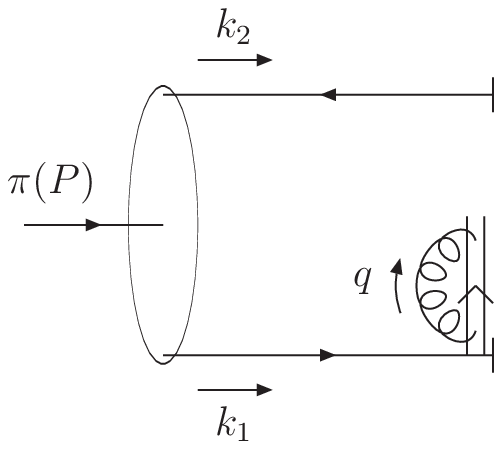}
\includegraphics[height=3cm]{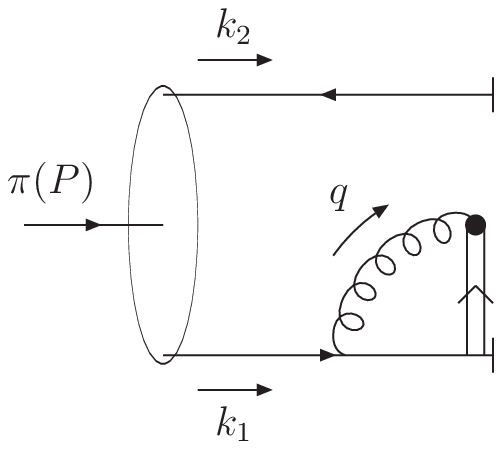}

\hspace{1.0cm} (a)\hspace{3.0cm} (b) \hspace{3.0cm} (c) \vskip
0.5cm

\includegraphics[height=3cm]{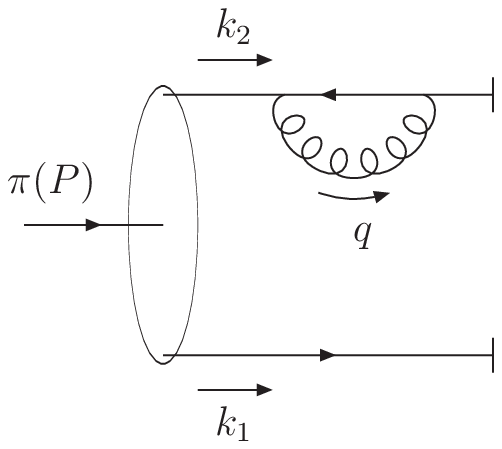}
\includegraphics[height=3cm]{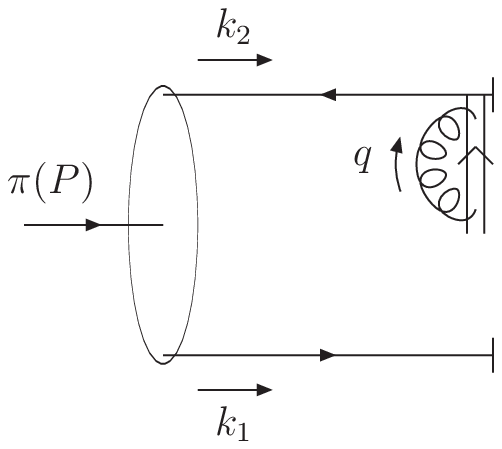}
\includegraphics[height=3cm]{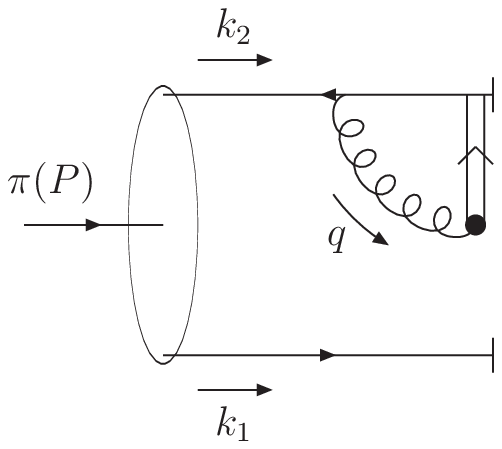}

\hspace{1.0cm} (d)\hspace{3.0cm} (e) \hspace{3.0cm} (f)

\caption{NLO diagrams for the pion wave function without the loop
momentum flowing through the hard kernel.} \label{fig3}
\end{center}
\end{figure}

We find that the mistakes made in the calculation of Fig.~\ref{fig2}
and Fig.~\ref{fig3} \cite{FMW08} arise from an improper application
of the contour integration. Take the simple loop integral associated
with Figs.~\ref{fig2}(d), \ref{fig3}(b) and \ref{fig3}(e) as an
example:
\begin{equation}
I = 16i\alpha g_s^2 \int \frac{ d^4q} {(2\pi)^4} \frac{1}
{(q^2-\lambda_L^2+ i\varepsilon)
[q^2-(1-\alpha)\lambda_L^2+ i\varepsilon]}\;,\label{4}
\end{equation}
where the gauge parameter $\alpha$ comes from the gluon propagator
in the covariant gauge \cite{FMW08}
\begin{equation}
  \frac{-i }{q^2 -\lambda_L^2+ i\varepsilon} \left [ g^{\mu\nu}
  - \alpha \frac{q^\mu q^\nu}{q^2 -(1-\alpha)\lambda_L^2+
    i\varepsilon}
\right]\;,\label{gp1}
\end{equation}
and the gluon mass $\lambda_L$ is introduced to isolate the
infrared (IR) pole. Applying the Feynman parametrization, it is
easy to obtain
\begin{equation}
I = \frac{4 \alpha\alpha_s}{\pi}\ln \frac{\lambda_L^2 } {\mu^2}
+{\rm UV\ pole}\;\label{3},
\end{equation}
where $\ln\lambda_L^2$ denotes the IR singularity
\cite{FMW08}, the explicit expression for the ultraviolet (UV) pole
is irrelevant here, and $\mu$ is the renormalization scale. However,
if employing the contour integration naively, Eq.~(\ref{4})
vanishes:
\begin{equation}
I = 16 i\alpha g_s^2 \int \frac{ d^4q} {(2\pi)^4}
\frac{1}{(2q^+q^--q_\perp^2 -\lambda_L^2+i\varepsilon)
[2q^+q^--q_\perp^2 -(1-\alpha)\lambda_L^2+i\varepsilon]}=0\;.\label{5}
\end{equation}
The poles in the above integrand are located in the same half plane,
either upper or lower, no matter whether one integrates over the $q^-$
or $q^+$ first. Therefore, one can close the contour in the other half
plane not containing the poles. Obviously, Eq.~(\ref{5}) contradicts
Eq.~(\ref{3}). How to understand this puzzle is the key to clarify the
conflict between \cite{FMW08} and \cite{NL07}.

\begin{figure}[t]
\begin{center}
\includegraphics[height=3cm]{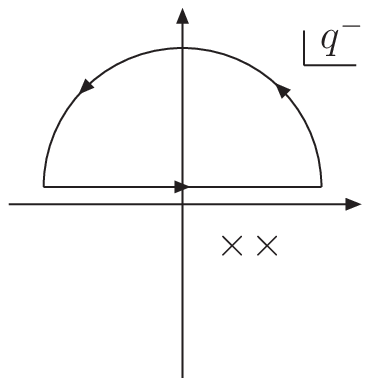}\hspace{1.0cm}
\includegraphics[height=3cm]{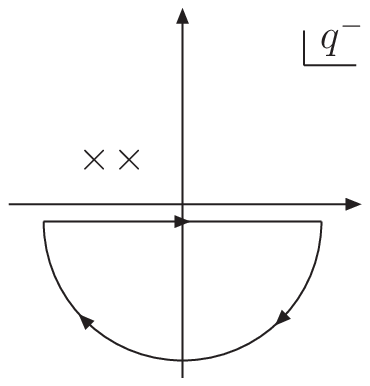}

\hspace{0.8cm} (a)\hspace{3.5cm} (b)

\caption{Contours for Eqs.~(\ref{5}) and (\ref{new1})
in the regions (a) $q^+>0$ and (b) $q^+<0$
.} \label{fig4}
\end{center}
\end{figure}

The above simple example illustrates that a naive application of the
contour integration in the light-cone coordinates may lead to a
false result. This is exactly the reason the authors of \cite{FMW08}
drew the wrong conclusion from Fig.~\ref{fig2}. When one closes the
contour in Eq.~(\ref{5}) by including the semicircle at infinity, it
has been implicitly assumed that the contribution from the
semicircle is negligible. However, this assumption does not hold as
$q^+\to 0$ \cite{M09}: the product
$q^+q^-$ in the denominator does not provide a suppression on the
semicircle of a large radius $|q^-|$ as $q^+\to 0$. In other words,
the poles $q^-=(q_\perp^2+\lambda_L^2)/(2q^+)$ and
$q^-=[q_\perp^2+(1-\alpha)\lambda_L^2]/(2q^+)$, also moving to
infinity as $q^+\to 0$, cannot be avoided by the contour in
Eq.~(\ref{5}). A safe way to proceed is to close the
contour of $q^-$ with a semicircle of a large but finite radius $R$
in the half plane without the poles. For example, the semicircle in
the upper half plane is considered for $q^+>0$ in Fig.~\ref{fig4}(a).
The integration over $q^-$ along the real axis
is thus equated to the integration over the semicircle, and Eq.~(\ref{5})
becomes
\begin{eqnarray}
I &=& 16i\alpha g_s^2 \lim_{R\to\infty}\int \frac{d^2q_\perp} {(2\pi)^4}
\left[i\int_\pi^0 d\theta\int_0^\infty dq^+ +i\int_{-\pi}^0
d\theta\int_{-\infty}^0
dq^+\right]\nonumber\\
& &\hspace{30mm}\times
\frac{Re^{i\theta}}{
(2q^+Re^{i\theta}-q_\perp^2 -\lambda_L^2+ i\varepsilon)
[2q^+Re^{i\theta}-q_\perp^2 -(1-\alpha)\lambda_L^2+ i\varepsilon]
}\;,\label{new1}
\end{eqnarray}
where the first and second terms in the above square brackets
correspond to Fig.~\ref{fig4}(a) and Fig.~\ref{fig4}(b),
respectively. Performing the integration over $q^+$ and
$\theta$, and then taking the $R\to\infty$ limit, we reproduce
Eq.~(\ref{3}). With the above prescription, the potential contribution
from the large semicircles at $q^+\to 0$ is taken into account, and a
consistent result is obtained in the contour integration.

Below we recalculate Fig.~\ref{fig2} (the NLO diagrams for the pion
wave function with the loop momentum flowing through the hard
kernel) following the aforementioned method, and demonstrate that
Figs.~\ref{fig2}(a), \ref{fig2}(b), and \ref{fig2}(c) are in fact
free of the gauge-dependent light-cone singularity
$\alpha\ln\lambda_L^2$. We also recalculate Fig.~\ref{fig3} (the NLO
diagrams for the pion wave function without the loop momentum
flowing through the hard kernel), and point out the erroneous
observation derived from Figs.~\ref{fig3}(c) and \ref{fig3}(f) in
\cite{FMW08}. Since the authors of \cite{FMW08} have evaluated the
NLO diagrams for the form factor using the Feynman parametrization,
these results are valid.

We start from Fig.~\ref{fig2}(b), whose gauge-dependent part is
written as
\cite{FMW08}
\begin{equation}
\phi_\alpha \vert_{2b}=16i \alpha g_s^2 \int \frac{d^4 q}{(2\pi)^4}
   \frac{ 2 (k_1^+-q^+) q^- -\vec k_{1\perp} \cdot  \vec q_\perp + q_\perp^2 }
      { [(k_1-q)^2 +i\varepsilon]
    ( q^2 + i\varepsilon )^2
}\delta (k^+ -(k_1^+-q^+)) \delta^2 (\vec k_\perp - (\vec k_{1\perp} -\vec q_{\perp}) )
      \;.\label{2b}
\end{equation}
The wave function $\phi_\alpha \vert_{2b}$ is convoluted with the LO
hard kernel:
\begin{equation}
\phi_\alpha \vert_{2b} \otimes H^{(0)}
  = \int_0^1 dx \int d^2 k_\perp \frac {1} { xQ^2 + k_\perp^2} \phi_\alpha \vert_{2b}\;,
\end{equation}
with the variable $x\equiv k^+/P^+$. Integrating the two
$\delta$-functions over $x$ and $k_\perp$, the LO hard kernel in
Eq.~(\ref{2}) appears. In the light-cone region the scaling law for
the components of $q$ is defined by $(q^+, q^-, q_\perp)\sim
(\delta^2, 1, \delta)$ with $\delta$ being a small parameter
\cite{FMW08}, according to which the first term $(k_1^+-q^+)q^-$ in
Eq.~(\ref{2b}) gives the leading contribution. The light-cone
singularity is regularized by introducing the gluon mass
$\lambda_L^2$ as shown in Eq.~(\ref{gp1}). Applying the contour
integration over $q^-$ for $0<q^+<k_1^+$ naively, we reproduce the
result in \cite{FMW08},
\begin{equation}
\phi_\alpha^{\rm FMW} \vert_{2b} \otimes H^{(0)}
   =  \frac{4\alpha \alpha_s}{P^+ \pi}
   \frac { \ln \lambda_L^2 } { x_0 Q^2 + k_{1\perp}^2} + {\rm finite\
   terms}\;.\label{2b1}
\end{equation}

\begin{figure}[t]
\begin{center}
\includegraphics[height=3cm]{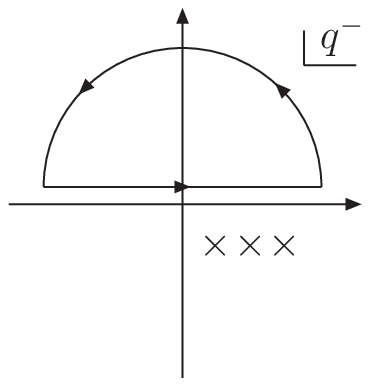}\hspace{1.0cm}
\includegraphics[height=3cm]{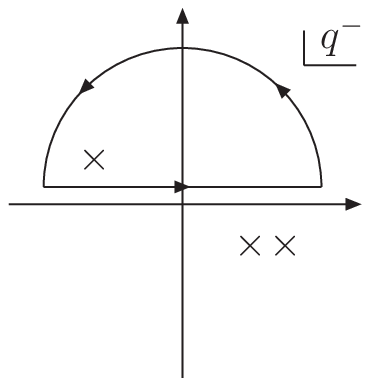}\hspace{1.0cm}
\includegraphics[height=3cm]{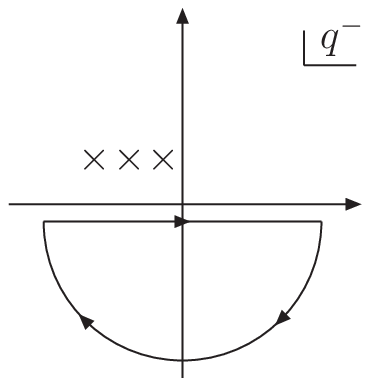}

\hspace{0.8cm} (a)
\hspace{3.3cm} (b)
\hspace{3.3cm} (c)

\caption{Contours for Eqs.~(\ref{2b}) and (\ref{2b-c})
in the regions
(a) $q^+>k_1^+$, (b) $0<q^+<k_1^+$, and (c) $q^+<0$.} \label{fig5}
\end{center}
\end{figure}

As explained above, we have to examine the potential contribution
from the large semicircles. For the range $q^+ >k_1^+$, all the
poles are located in the lower half plane, so we close the contour
of $q^-$ with a large semicircle in the upper half plane shown in
Fig.~\ref{fig5}(a). The contribution from this semicircle remains
negligible in the limit $q^+\to k_1^+$, because of the suppression
factor $(k_1^+-q^+)$ in the numerator of Eq.~(\ref{2b}). For
$q^+<0$, all the poles are located in the upper half plane, so we
close the contour with a large semicircle in the lower plane shown
in Fig.~\ref{fig5}(c). In this case, the semicircle may contribute
as $q^+\to 0$, and we equate the integral over $q^-$ to the
contribution from the semicircle of a radius $R$. For $0<q^+<k_1^+$,
the single pole $q^-=(|\vec k_{1\perp}-\vec q_\perp|^2-i\varepsilon)
/[2(q^+-k_1^+)]$ is located in the upper half plane and the double
pole $q^-=(q_\perp^2-i\varepsilon)/(2q^+)$ in the lower half plane.
We close the contour of $q^-$ through the upper half plane with a
semicircle shown in Fig.~\ref{fig5}(b), which picks up the residue
from the single pole, giving Eq.~(\ref{2b1}). The contribution from
the semicircle of a radius $R$ needs to be subtracted, since it does
not vanish at infinity as $q^+\to 0$.

Therefore, Fig.~\ref{fig2}(b), as convoluted with the LO hard
kernel, contains the additional contribution from the two
semicircles, one for $0<q^+<k_1^+$ and another for $q^+<0$,
\begin{eqnarray}
\phi'_\alpha \vert_{2b} \otimes H^{(0)}&=& \frac{16i\alpha
g_s^2}{P^+} \lim_{R\to\infty}\int \frac{d^2q_\perp} {(2\pi)^4}
\left[i\int_\pi^0 d\theta\int_0^{k_1^+} dq^++i\int_{-\pi}^0
d\theta\int_{-\infty}^0
dq^+\right]\nonumber\\
& &\times
\frac{2(k_1^+-q^+)(Re^{i\theta})^2}{(2q^+Re^{i\theta}-q_\perp^2-\lambda_L^2)
[2q^+Re^{i\theta}-q_\perp^2-(1-\alpha)\lambda_L^2]
[2(q^+-k_1^+)Re^{i\theta}-|{\vec
k}_{1\perp}-{\vec q}_\perp|^2]}\nonumber\\
& &\times \frac{1}{(k_1^+-q^+)Q^2/P^++|\vec k_{1\perp}- \vec
q_\perp|^2}\;. \label{2b-c}
\end{eqnarray}
Working out the integration over $q^+$, $\theta$ and $q_\perp$, and
then taking the $R\to\infty$ limit, we obtain
\begin{eqnarray}
\phi'_\alpha \vert_{2b} \otimes H^{(0)}= -\frac{4\alpha
\alpha_s}{P^+ \pi}\frac { \ln \lambda_L^2 } { x_0 Q^2 +
k_{1\perp}^2} + {\rm finite\
   terms}\;,\label{2b2}
\end{eqnarray}
which cancels the IR pole in Eq.~(\ref{2b1}) exactly, implying that
Fig.~\ref{fig2}(b) is free of the gauge-dependent light-cone
singularity. Note that the limit $R\to\infty$ must be taken after
the $q^+$ integration. Then the contribution from the semicircles at
$q^+=0$ can be evaluated correctly. The same conclusion applies to
Fig.~\ref{fig2}(c).

We have explained the absence of the light-cone singularity in
Fig.~\ref{fig2}(b) through the contour integration over $q^-$ by
including the semicircle contribution. Each of the residue
contribution in Eq.~(\ref{2b1}) and the semicircle contribution in
Eq.~(\ref{2b2}) has the light-cone singularity, but they cancel each
other exactly. It is thus understood why the authors in \cite{FMW08}
got a fake light-cone singularity from Fig.~\ref{fig2}(b): they have
missed the contribution from the large semicircles. Their wrong
result in Eq.~(\ref{2b1}) has an origin similar to that of
Eq.~(\ref{5}). Performing the integration over $q^+$ first, we
obtain the same result as in the case starting with the $q^-$
integration. In the contour integration over $q^+$, the semicircle
contribution must be handled appropriately as $q^-\to 0$. For the
range $q^- >0$, we equate the integral over $q^+$ to the
contribution from the semicircle of a radius $R$ in the upper half
plane. For $q^-<0$, we close the contour of $q^+$ through the lower
half plane with a semicircle of a radius $R$, which picks up the
residue from the pole $q^+=(k_1^+Q^2+P^+|\vec k_{1\perp}-\vec
q_\perp|^2 -i\varepsilon)/Q^2$. It is found that neither the residue
contribution nor the semicircle contribution contains the light-cone
singularity, because the integral in Eq.~(\ref{2b}) may be singular
in the region $(q^+,q^-,q_\perp)\sim(\delta^2,1,\delta)$, but not as
$q^-\to 0$.

If applying the contour integration naively to Fig.~\ref{fig3}(c),
\begin{equation}
\phi_\alpha \vert_{3c} \otimes H^{(0)}
=-\frac{16i \alpha g_s^2}{P^+} \int \frac{d^4 q}{(2\pi)^4}
   \frac{ 2 (k_1^+-q^+) q^- -\vec k_{1\perp} \cdot  \vec q_\perp + q_\perp^2 }
      { [(k_1-q)^2 +i\varepsilon]
    ( q^2 + i\varepsilon )^2
   ( x_0Q^2 + k_{1\perp}^2) }\;,
\end{equation}
the mistake made in \cite{FMW08} is more obvious. The term
$(k_1^+-q^+) q^-$ gives a light-cone singularity from the region
$(q^+,q^-,q_\perp)\sim(\delta^2,1,\delta)$ in the naive contour
integration, which is the same as in Eq.~(\ref{2b1}). Since the loop
momentum does not flow through the hard kernel, the term
$(k_1^+-q^+) q^-$ also generates a UV divergence from the region
$(q^+,q^-,q_\perp)\sim (1,\Lambda^2,\Lambda)$ with the scale
$\Lambda\to\infty$ in the naive contour integration. However,
according to the covariance argument \cite{FMW08}, the corresponding
loop integral, proportional to $q^-$, should vanish like $k_1^-=0$.
To overcome this apparent contradiction, the authors in \cite{FMW08}
dropped the gluon mass $\lambda_L$, adopted the dimensional
regularization, misinterpreted the UV divergence as another
light-cone singularity, and made the UV divergence and the
light-cone singularity cancel each other. The momentum configuration
$q\sim (1,\Lambda^2,\Lambda)$ should give a UV divergence, since it
must be regularized with the number of dimensions $n<4$ in the
dimensional regularization, and cannot be regularized by the gluon
mass $\lambda_L$. Hence, we have difficulty to understand their 
argument \cite{FMW08} that a gluon with
infinite invariant mass $q^2\sim \Lambda^2\to\infty$ produces the
light-cone (infrared) singularity.
The fact is that Fig.~\ref{fig3}(c) has neither UV divergence nor
light-cone singularity, after carefully including the contribution
from the semicircles.

Our result is natural from the viewpoint of the Ward identity.
Contracting the loop momentum $q$ to the vertex on the internal
quark line (see the numerator of the gauge-dependent term in
Eq.~(\ref{gp1})),
\begin{eqnarray}
\frac{\not p-\not k_1}{(p-k_1)^2}\not q \frac{\not
p-\not k_1+\not q}{(p-k_1+q)^2}=\frac{\not
p-\not k_1}{(p-k_1)^2}-\frac{\not p-\not
k_1+\not q}{(p-k_1+q)^2}\;,
\end{eqnarray}
the first (second) term leads to the LO hard kernel associated with
Fig.~\ref{fig3}(c) (Fig.~\ref{fig2}(b)). The diagram Fig.~4c in
\cite{FMW08} for the form factor is then factorized into
convolutions of the LO hard kernel with Figs.~\ref{fig2}(b) and
\ref{fig3}(c). If Fig.~4c in \cite{FMW08} and Fig.~\ref{fig3}(c) do
not contain the light-cone singularity as observed in \cite{FMW08},
Fig.~\ref{fig2}(b) should not either. Another support to our result
comes from the following simple observation. In the light-cone
region the $q^+$ and $q_\perp$ dependence in the LO hard kernel is
negligible. Then Figs.~\ref{fig2}(b) and \ref{fig3}(c) involve
exactly the same loop integral, namely, the same behavior in the
light-cone region. If Fig.~\ref{fig3}(c) does not produce the
light-cone singularity as found in \cite{FMW08}, Fig.~\ref{fig2}(b)
should not either.

We then turn to Fig.~\ref{fig2}(a), which appears as a consequence
of factorizing the box diagram Fig.~4a of \cite{FMW08} for the form
factor. Again, Fig.~\ref{fig2}(a) should not contain the
gauge-dependent light-cone singularity, because Fig.~4a of
\cite{FMW08} does not. The explicit expression for the leading
contribution from Fig.~\ref{fig2}(a) in the light-cone region is
given by
\begin{eqnarray}
\phi_\alpha \vert_{2a}&=&-64i \alpha g_s^2 \int \frac{d^4
q}{(2\pi)^4}
   \frac{(k_1^+-q^+) (k_2^++q^+)(q^-)^2}
      { [(k_1-q)^2 +i\varepsilon][(k_2+q)^2 +i\varepsilon]
      (q^2 +i\varepsilon)^2}\nonumber\\
& &\hspace{2.0cm}\times\delta (k^+ -(k_1^+-q^+)) \delta^2 (\vec k_\perp -
(\vec k_{1\perp} -\vec q_\perp) )
      \;,\label{2a}
\end{eqnarray}
where the numerator $(q^-)^2$ arises from $q^\mu q^\nu$ in the gluon
propagator for $\mu=\nu=-$. Integrating over $q^-$ first, we have
two poles, $q^-=(|\vec k_{1\perp}-\vec
q_\perp|^2-i\varepsilon)/[2(q^+-k_1^+)]$ for $0<q^+<k_1^+$ and
$q^-=(|\vec k_{1\perp}-\vec q_\perp|^2-i\varepsilon)/[2(q^++k_2^+)]$
for $-k_2^+<q^+<0$. Closing the contour in the upper half plane for
the former, and in the lower half plane for the latter naively, the
result in \cite{FMW08} is reproduced,
\begin{eqnarray}
\phi_\alpha^{\rm FMW} \vert_{2a} \otimes H^{(0)} &=&
   -\frac{4\alpha \alpha_s}{P^+ \pi}
   \frac {\ln \lambda_L^2 } { x_0 Q^2 + k_{1\perp}^2} + {\rm finite\
   terms}\;,\label{2a1}
\end{eqnarray}
which is also a fake singularity.

Similarly, the contribution from the two semicircles of a radius
$R$, which were added to form the closed contours mentioned above,
is written as
\begin{eqnarray}
\phi'_\alpha \vert_{2a} \otimes H^{(0)}&=&
-\frac{256i\pi\alpha\alpha_s}{P^+} \lim_{R\to\infty} \int
\frac{d^2q_\perp} {(2\pi)^4} \left[i\int^0_{\pi}
d\theta\int_0^{k_1^+} dq^++i\int_{-\pi}^0 d\theta\int_{-k_2^+}^0
dq^+ \right]
\nonumber\\
&&\hspace{10mm}\times \frac{(k_1^+-q^+)(k_2^++q^+)(Re^{i\theta})^3}
{\left[-2(k_1^+-q^+)Re^{i\theta}-|\vec k_{1\perp}-\vec
q_\perp|^2+i\varepsilon\right]
 \left[2(k_2^++q^+)Re^{i\theta}-|\vec k_{1\perp}-\vec q_\perp|^2+i\varepsilon\right]}
\nonumber\\
&&\hspace{10mm}\times \frac{1} {
(2q^+Re^{i\theta}-q_\perp^2-\lambda_L^2+i\varepsilon)
  (2q^+Re^{i\theta}-q_\perp^2-(1-\alpha)\lambda_L^2+i\varepsilon)}
\nonumber\\
&&\hspace{10mm} \times \frac{1}{ (k_1^+-q^+)Q^2/P^++|\vec
k_{1\perp}-\vec q_\perp|^2-i\varepsilon}\;.\label{2as}
\end{eqnarray}
The rest of the procedure is straightforward, leading to
\begin{eqnarray}
\phi'_\alpha \vert_{2a} \otimes H^{(0)}= \frac{4\alpha \alpha_s}{P^+
\pi}\frac { \ln \lambda_L^2 } { x_0 Q^2 + k_{1\perp}^2} + {\rm
finite\
   terms}\;.\label{2a2}
\end{eqnarray}
The sum of Eqs.~(\ref{2a1}) and (\ref{2a2}), i.e.,
Fig.~\ref{fig2}(a), is free of the gauge-dependent light-cone
singularity, contrary to the observation in \cite{FMW08}. In
summary, all the diagrams in Figs.~\ref{fig2} and \ref{fig3}, except
those with the gluons attaching only the Wilson lines
(Figs.~\ref{fig2}(d), \ref{fig3}(b), and \ref{fig3}(e)), do not
generate the gauge-dependent IR singularity.

We then comment on the diagrams Figs.~\ref{fig2}(d), \ref{fig3}(b),
and \ref{fig3}(e). These diagrams do not appear in the factorization
of collinear gluons \cite{Collins03,L1,NL2} from the pion transition
form factor. It has been shown that their sum is IR finite
\cite{FMW08}. This must be the case, since all the IR divergences in
the form factor diagrams, which arise from the collinear region with
the scaling law $(q^+, q^-, q_\perp)\sim (1,\delta^2, \delta)$, have
been absorbed into the other diagrams in Figs.~\ref{fig2} and
\ref{fig3} \cite{NL07}. However, in view of a gauge invariant
definition for the pion wave function, Figs.~\ref{fig2}(d),
\ref{fig3}(b), and \ref{fig3}(e) should be included. A resolution is
to invoke a soft subtraction factor in the denominator
\cite{Collins03,FMW08}:
\begin{eqnarray}
\phi(x_0;x,k_\perp)=\int\frac{dy^-}{2\pi
i}\frac{d^2y_\perp}{(2\pi)^2}e^{-i(1-x)P^+ y^--i{\vec k}_\perp\cdot
{\vec y}_\perp}\frac{\langle 0|{\bar q}(y)
W_y(n)^{\dag}I_{n;y,0}W_0(n)\gamma^+\gamma_5 q(0) |q(k_1)\bar
q(k_2)\rangle} {\langle
0|W_y(n)^{\dag}W_y(u)I_{n;y,0}I_{u;y,0}^{\dag}
W_0(n)W_0(u)^{\dag}|0\rangle}\;, \label{phim2}
\end{eqnarray}
which removes Figs.~\ref{fig2}(d), \ref{fig3}(b), and \ref{fig3}(e)
in a gauge invariant way. In the above definition, $y=(0,y^-,{\vec
y}_\perp)$ is the coordinate of the anti-quark field $\bar q$ , $n$
with $n^2\not= 0$ the direction of the Wilson line, and $|q(k_1)\bar
q(k_2)\rangle$ the leading Fock state of the pion. The factor
$W_y(n)$ denotes the Wilson line operator
\begin{eqnarray}
\label{eq:WL.def} W_y(n) = P \exp\left[-ig_s \int_0^\infty d\lambda
n\cdot A(y+\lambda n)\right]\;.
\end{eqnarray}
The two Wilson lines $W_y(n)$ and $W_0(n)$ must be connected by a
link $I_{n;y,0}$ at infinity \cite{NL2,BJY}.

The subtraction factor in the denominator generates the soft
diagrams similar to Figs.~\ref{fig2} and \ref{fig3}, but with the
fermion lines being replaced by the Wilson lines in the direction
$u$. Besides Figs.~\ref{fig2}(d), \ref{fig3}(b), and \ref{fig3}(e),
additional soft diagrams, such as the vertex corrections with the
gluons attaching the Wilson lines in the directions $n$ and $u$, are
introduced at the same time. To ensure that the soft subtraction
does not change the collinear structure of the numerator, $u$ should
not lie on the light cone, namely, $u^2\not= 0$. Other than this
requirement, the direction of $u$ is completely arbitrary. Hence,
the subtraction of Figs.~\ref{fig2}(d), \ref{fig3}(b), and
\ref{fig3}(e) will result in an arbitrary UV pole for the pion wave
function. One can then take advantage of this arbitrariness,
adjusting $u$ so that the UV pole of the above vertex corrections
cancels the UV pole of the self-energy corrections to the Wilson
lines along $u$. Below we demonstrate this cancellation explicitly.
Because the subtraction factor is gauge invariant, we evaluate the
soft diagrams in the Feynman gauge, i.e., consider only the
$g^{\mu\nu}$ tensor for the gluon propagator. Then the former is
given by \cite{LL04}
\begin{eqnarray}
-\frac{\alpha_sC_F}{4\pi}\frac{u\cdot n}{\sqrt{(u\cdot n)^2-u^2n^2}}
\ln\frac{\sqrt{(u\cdot n)^2-u^2n^2}+u\cdot n}
{\sqrt{(u\cdot n)^2-u^2n^2}-u\cdot n}\left(\frac{1}{\epsilon}
+\ln\frac{4\pi\mu^2}{\lambda_L^2e^{\gamma_E}}\right)\;,
\end{eqnarray}
and the latter by
\begin{eqnarray}
\frac{\alpha_sC_F}{4\pi}\left(\frac{1}{\epsilon}
+\ln\frac{4\pi\mu^2}{\lambda_L^2e^{\gamma_E}}\right)\;.
\end{eqnarray}
It is easy to find that the above two expressions cancel exactly for
$u^2n^2<0$ and $u\cdot n/\sqrt{|u^2n^2|}\approx 0.85$. It is also
trivial to show that the remaining soft diagrams with the loop
momentum flowing through the hard kernel cancel, when the above
conditions for $u$ and $n$ are satisfied. Eventually, we just need
to calculate the seven effective diagrams in Fig.~2 of \cite{NL07},
i.e., Figs.~\ref{fig2}(a)-\ref{fig2}(c), and Figs.~\ref{fig3}(a),
\ref{fig3}(c), \ref{fig3}(d) and \ref{fig3}(f) in this comment,
in the Feynman gauge at one-loop level.

After clarifying the conflict, the all-order proof of the
leading-power $k_T$ factorization theorem for the pion transition
form factor in Sec.~III of \cite{NL07} follows straightforwardly.
Below we mention the points of the proof briefly. Since the
light-cone singularity does not exist, the substitution
for the metric tensor of the gluon propagator in the covariant
gauge, Eq.~(47) of \cite{NL07}, works for extracting all IR
(collinear) divergences. We then employ the derivative in Eq.~(50)
of \cite{NL07} to collect the gauge-dependent terms arising from the
off-shell partons. The derivative of the soft subtraction factor
vanishes, since it is gauge invariant. The derivative of the pion
wave function is then related to the derivative of the numerator in
Eq.~(\ref{phim2}), which, after applying the Ward identity, gives
Eq.~(53) in \cite{NL07}. Combining Eqs.~(52) and (53) in
\cite{NL07}, the $k_T$ factorization theorem for the pion transition
form factor is proved by induction.

In conclusion, the only necessary revision for \cite{NL07} is to
replace the definition for the $k_T$-dependent pion wave function by
Eq.~(\ref{phim2}) with an appropriate vector $u$. Except this
modification, all the results in \cite{NL07} are valid. The
gauge-dependent light-cone singularity found in \cite{FMW08} does
not exist; their result is attributed to a careless application of
the contour integration in the light-cone coordinates. The
correct prescription is to keep the contribution from the
semicircles of a finite radius first, work out the loop
integration, and then move the semicircles to infinity.
We have confirmed that the all-order proof of the $k_T$
factorization theorem for the pion transition form factor holds: the
gauge dependence, arising from the off-shell external partons,
cancels between the full QCD diagrams for the form factor and the
effective diagrams for the pion wave function. Therefore, the $k_T$
factorization produces a gauge invariant and IR finite hard kernel,
and provides a solid basis for the PQCD approach to exclusive $B$
meson decays.

\vskip 0.3cm We thank J.P. Ma and Q. Wang for useful discussions.
The work was supported in part by the National Science Council of
R.O.C. under Grant No. NSC-95-2112-M-050-MY3, by the National
Center for Theoretical Sciences of R.O.C., and by Kavli Institute
for Theoretical Physics, China.


\begin{thebibliography}{99}

\bibitem{FMW08} F. Feng, J.P. Ma, and Q. Wang, arXiv:0807.0296
[hep-ph].
\bibitem{LS} H-n. Li and G. Sterman, Nucl. Phys. {\bf B381}, 129 (1992).
\bibitem{LY1} H-n. Li and H.L. Yu, Phys. Rev. Lett. {\bf 74}, 4388 (1995);
Phys. Lett. B {\bf 353}, 301 (1995); Phys. Rev. D {\bf 53}, 2480
(1996).
\bibitem{KLS} Y.Y. Keum, H-n. Li, and A.I. Sanda,
Phys. Lett. B {\bf 504}, 6 (2001); Phys. Rev. D {\bf 63}, 054008
(2001); Y.Y. Keum and H-n. Li, Phys. Rev. {\bf D63}, 074006
(2001).
\bibitem{LUY} C. D. L\"{u}, K. Ukai, and M. Z. Yang, Phys. Rev. D {\bf 63},
074009 (2001).
\bibitem{NL07} S. Nandi and H-n. Li, Phys. Rev. D {\bf 76},
034008 (2007).
\bibitem{M09} J.P. Ma, private communication.
\bibitem{Collins03} J.C. Collins, Adv. Ser. Direct. High Energy Phys.
{\bf 5}, 573 (1989).
\bibitem{L1} H-n. Li, Phys. Rev. D {\bf 64}, 014019 (2001); M. Nagashima
and H-n. Li, Eur. Phys. J. C {\bf 40}, 395 (2005).
\bibitem{NL2} M. Nagashima and H-n. Li, Phys. Rev. D {\bf 67},
034001 (2003).
\bibitem{BJY} X. Ji, and F. Yuan, Phys. Lett. B {\bf 543}, 66 (2002);
A.V. Belitsky, X. Ji, and F. Yuan, Nucl. Phys. {\bf B656}, 165
(2003).
\bibitem{LL04} H-n. Li and H.S. Liao, Phys. Rev. D {\bf 70}, 074030 (2004).





\end{thebibliography}
\end{document}